\documentclass[%
 reprint,
superscriptaddress,
 amsmath,amssymb,
 aps,
 pre,
]{revtex4-1}

\usepackage{graphicx}
\usepackage{dcolumn}
\usepackage{bm}

\let\oldhat\hat
\renewcommand{\vec}[1]{\mathbf{#1}}
\renewcommand{\hat}[1]{\oldhat{\mathbf{#1}}}

\begin{document}

\preprint{APS/123-QED}

\title{Forces in inhomogeneous open active-particle systems}

\author{Nitzan Razin}
 \affiliation{Department of Chemical Physics, Weizmann Institute of Science, Rehovot 76100, Israel}
\author{Raphael Voituriez}%
\affiliation{Laboratoire Jean Perrin and Laboratoire de Physique Th\'eorique de la Mati\`ere Condens\'ee, CNRS/Universit\'e Pierre et Marie Curie, 75005 Paris, France}%
\author{Jens Elgeti}
\affiliation{Theoretical Soft Matter and Biophysics, Institute of Complex Systems and Institute for Advanced Simulation, Forschungszentrum J\"ulich, D-52425 J\"ulich, Germany}%

\author{Nir S. Gov}
\affiliation{Department of Chemical Physics, Weizmann Institute of Science, Rehovot 76100, Israel}%

\date{\today}

\begin{abstract}
We study the force that non-interacting point-like active particles apply to a symmetric inert object in the presence of a gradient of activity and particle sources and sinks. We consider two simple patterns of sources and sinks that are common in biological systems. We analytically solve a one dimensional model designed to emulate higher dimensional systems, and study a two dimensional model by numerical simulations. We specify when the particle flux due to the creation and annihilation of particles can act to smooth the density profile that is induced by a gradient in the velocity of the active particles, and find the net resultant force due to both the gradient in activity and the particle flux. These results are compared qualitatively to observations of nuclear motion inside the oocyte, that is driven by a gradient in activity of actin-coated vesicles.
\end{abstract}

\maketitle

\section{Introduction}
The pressure exerted by active particles on surfaces and objects has recently attracted much attention \cite{Solon2015,Solon2015PRL,Nikola2016,Marchetti201634,Takatori2014,Yang2014,Yan2015,Ezhilan2015,Fily2014,Winkler2015,Mallory2014,Junot2017,Razin2017}. While generally, the pressure that dry active matter exerts on flat surfaces depends on the details of the particle-surface interaction, for spherical active particles with uniform motion parameters it is a state function \cite{Solon2015}. The particle density and also the pressure the particles exert on a curved surface depend on the local and even global surface curvature \cite{Nikola2016, Fily2014,Fily2015,Harder2014,Mallory2014b,Smallenburg2015}. For spherical particles, when a gradient in the properties of the particle motion - the speed and persistence time - exists, in the small persistence length limit the force on an immersed passive object is simply an integral over the object's surface of a local pressure which depends on the particle properties (but not, for example, on the shape of the surface). This force tends to push the object towards small persistence length regions. In two dimensions or higher, as the persistence length grows, additional contributions to the force, which do depend on the object's geometry, emerge \cite{Razin2017}.


Recent experiments \cite{Almonacid2015} showed that the nucleus of the mouse oocyte moves from the cortex to the center of the cell due to the active random motion of actin-coated vesicles. The density and velocity of the vesicles were measured as a function of the distance to the center of the spherically symmetric cell. While the density of the vesicles was found to be uniform, their measured velocity increases from the cell center towards the cortex. Note that the measured velocity $v_{mes}$ does not equal the momentary velocity of the vesicles. The measured velocity is equal to a position difference over the sample time, and it therefore depends on the time window used to determine the particle displacements. The biological mechanisms that give rise to the gradient in activity (or measured vesicle velocity), are not known at present. In this work we use this biological system as motivation to study the physics of systems of dry active particles with activity gradients. While in \cite{Razin2017} we studied the force the particles apply to objects in a closed system, we study here the possibility of particle creation/annihilation. Such active systems, with particle turnover, have not been previously studied, as far as we know.

We model the dilute fluid of vesicles, pulled by myosin Vb molecular motors walking on actin filaments, as non-interacting point-like active particles moving through a viscous fluid (overdamped dynamics).
The vesicles perform a locally random active motion, which is spatially nonuniform. To identify the fundamental underlying principles, we use a minimalistic modeling approach, only taking the key ingredients - particle speed and persistence - into account. We neglect further mechanisms, that would only deter from the minimalistic approach. In particular it is uncertain to which degree Hydrodynamic forces matter, as vesicles can exchange momentum with both the fluid and the actin network. 


In a previous work \cite{Razin2017}, we showed that in a closed system, it is possible to have a uniform particle density but a spatially varying measured velocity $v_{mes}$, and a force pushing an inert object, only by having a spatially dependent persistence time and a uniform speed, since the density is inversely proportional to the speed \cite{Schnitzer1993, Cates2015, Razin2017}. In this paper, we explore the possibility that the persistence time of the particles is uniform, while their speed increases from the center to the edge of the system. Since the density profile becomes highly non-uniform in such a system, we explore if particle turnover can restore a uniform density profile. However, the presence of particle turnover also introduces particle flows in the system, and these fluxes can induce a force on the passive object that conflicts with the force due to the activity gradient. We test whether it is possible to achieve an approximately uniform density while maintaining a force towards the center, by using particle sources and sinks, in two patterns that are common in biological systems.
We find a parameter regime where it is possible, and discuss whether it could occur in the oocyte, where vesicle trajectories were imaged in a two dimensional cross section of the three dimensional cell \cite{Almonacid2015}, and thus the rates and distribution of vesicle creation and annihilation could not be directly measured.

In order to study this problem, we construct a 1D model that emulates higher dimensional systems and is analytically solvable. In addition, we study 2D simulations with a similar geometry to the 3D oocyte experiment and show that they give results that are similar to the 1D model. 

\section{One dimensional model}

\begin{figure}[h]
\includegraphics[width=1\linewidth]{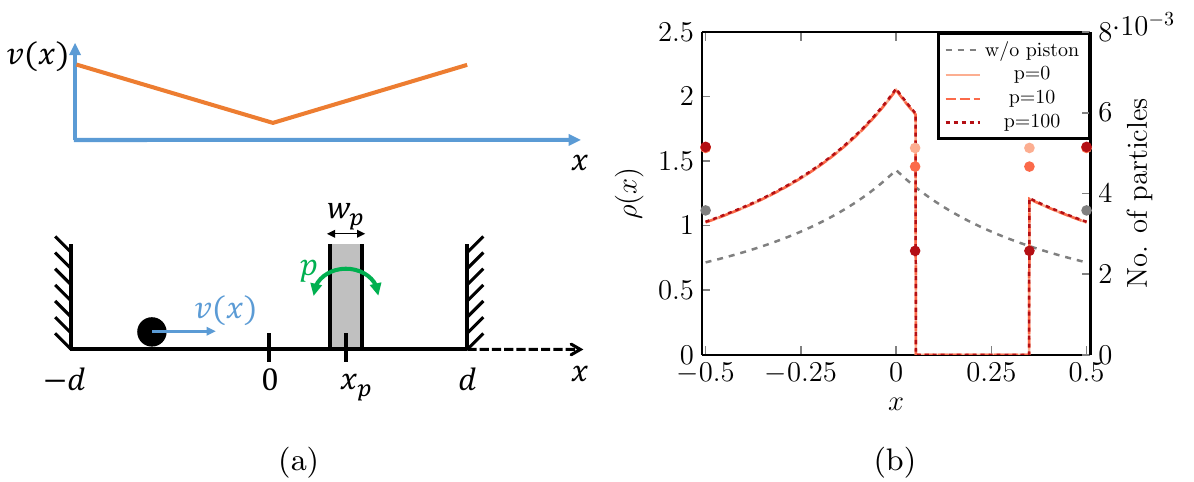}
  \caption[]{(a) Sketch of the 1D system with a permeable piston, and the particle velocity throughout the system (b) For a system with constant $\alpha$ and $v(x)=|x|+0.5$ ($N=1$): the particle density (lines) and number of particles accumulated on edges (dots), with (color) and without (gray) the piston (Eq.~\ref{eq:permeable_piston_sol_left}, \ref{eq:permeable_piston_sol_right}). Different colors specify different rates of particle passing through the piston $p$ ($x_p=0.2$, $w_p=0.3$). The bulk densities are indistinguishable for the different $p$ values, only the edge accumulations vary substantially. All length scales are measured in units of $x_0\equiv2d$, all time scales are measured in units of $\tau_0 \equiv 200/\alpha$.}
\label{fig:sys_plot_v_x_rho_p}
\end{figure}

We begin with a one dimensional model (sketched in Fig.~\ref{fig:sys_plot_v_x_rho_p}a) for point-like active particles obeying run-and-tumble dynamics, which are confined by hard walls in the domain $-d\le x \le d$. The motion of the particles is characterized by the tumble rate $\alpha$, which we assume to be constant, and speed $v(x)$, which is allowed to be a function of the particle position \cite{Schnitzer1993,Tailleur2008,Tailleur2009}. We assume that the walls have no effect on the orientation of the particles. We neglect thermal diffusion and interactions between the particles, for simplicity and since they are negligible in the biological system of interest \cite{Almonacid2015}.

The bulk density of particles in such a system, in the absence of fluxes, is $\rho(x)\propto 1/v(x)$ \cite{Schnitzer1993,Cates2015,Razin2017} (Fig.~\ref{fig:sys_plot_v_x_rho_p}b). Since a spatial variation in the bulk density is not observed for the biological system that motivates our study \cite{Almonacid2015}, we wish to explore processes that may allow us to decouple $\rho(x)$ from $v(x)$. A possible mechanism is to use particle sources and sinks, which is natural in biological contexts where objects such as vesicles are formed and have finite lifetimes. Such processes result in a steady state density that balances between the $v(x)$ and flux effects.


We study the motion of an inert object inside the system, representing for example the motion of the nucleus inside the oocyte. We will assume that the motion of the object inside the system is slow enough that the active particles attain their steady state density at all times. Furthermore, if the object's motion obeys overdamped dynamics, its velocity is proportional to the force applied to it. Thus we can calculate the mean steady state force on an object held at a fixed position and obtain the local velocity of a slow moving object. In 1D, such an object is a piston. However, a hard piston divides the system into disconnected parts. In order to avoid this pathology, which is not present in higher dimensions (where particles can move around an object), we make the piston permeable: a particle accumulated at one edge of the piston has probability $p$ per unit time to cross to the other side.

In Appendix \ref{appendix:equal_accum} we show that in the absence of particle fluxes, as long as the crossing rate from side to side is symmetric, the numbers of particles accumulated on each of the edges of the piston is equal. Thus this piston permeability destroys the mechanism of creating a force on the piston by unequal accumulations on its two edges, which occurs when a gradient in $\alpha$ exists \cite{Razin2017}. Nonetheless, since we study here the motion of the piston due to a gradient in $v$, where in the absence of flux, the particle accumulation on all surfaces is equal due to the constant $\alpha$ \cite{Razin2017}, we will use a symmetric permeable piston. We calculate in Appendix \ref{appendix:1D_sym_pp_sol} the steady state density in a closed system with a symmetric permeable piston. The result is plotted in Fig.~\ref{fig:sys_plot_v_x_rho_p}b for $v(x)=a|x|+b$ and $\alpha(x)=const$. The plot shows that increasing $p$ decreases the (equal) accumulation on the piston edges.

\subsection{Uniform particle creation, annihilation at the cortex}
\begin{figure}[h]
\includegraphics[width=1\linewidth]{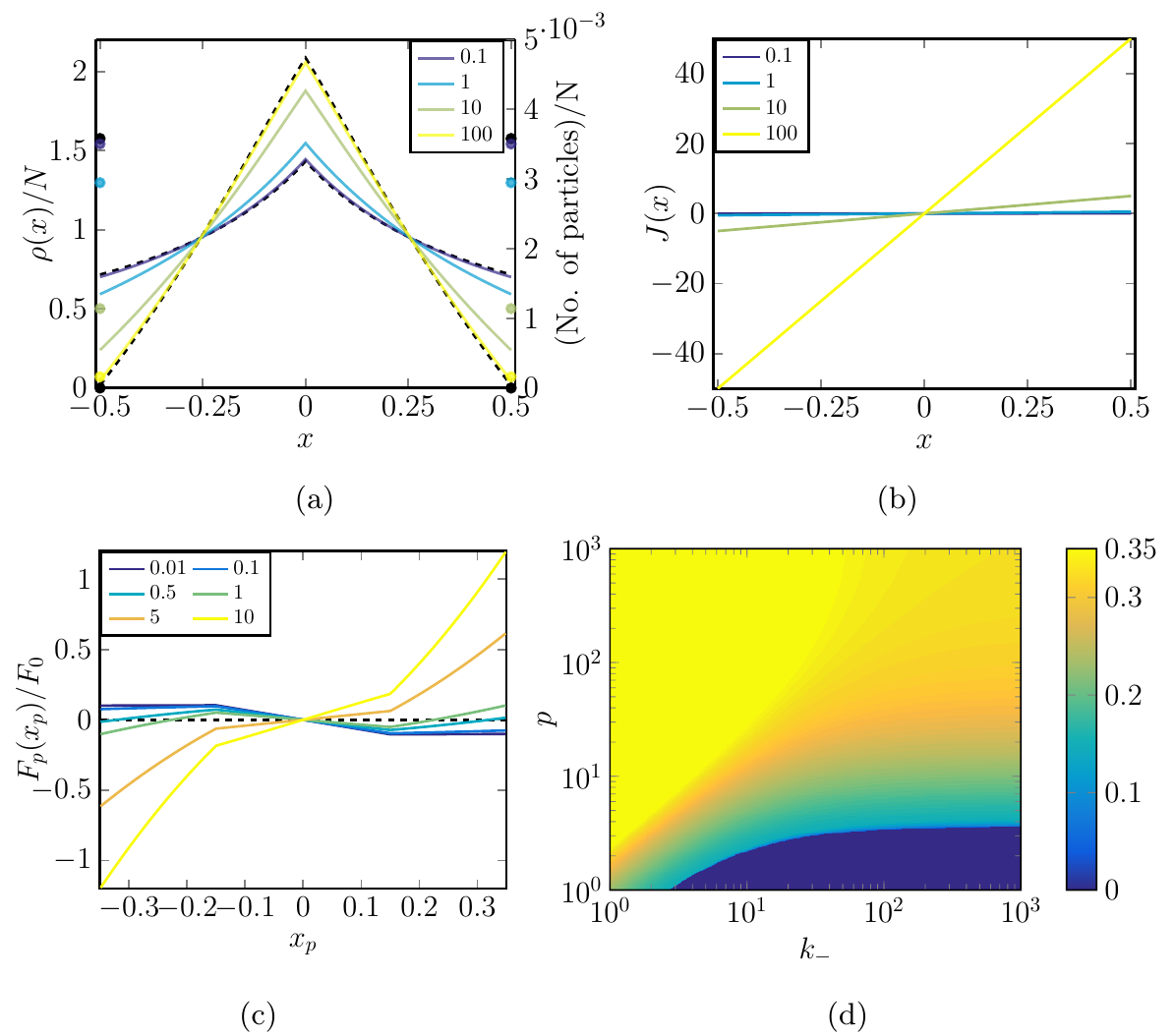}
  \caption[]{For a system with a constant $\alpha$ and $v(x)=|x|+0.5$: (a) The particle density (lines) and the number of particles accumulated on edges (dots) (Eq.~\ref{eq:FP_bulk_c_wall_a_v_x_Dr_x_D_0_sol_linear_v}), both divided by the total number of particles, for varying values of $k_-$. The limits of $k_+,k_-\to 0,\infty$, keeping $8dk_+/k_-=1$, are shown in dashed black lines and black dots. (b) The particle current $J(x)=k_+ x$ for varying values of $k_+$. ((a) and (b) - for a system without a piston) (c) The force on the piston for varying $k_-$ values ($w_p=0.3$, $p=1$, $8dk_+=k_-$) (d) The maximal position $0\le x_p\le d-w_p/2$ where the force on the piston is towards the center: $F_p(x_p)\le 0$ ($w_p=0.3$). As the particle annihilation rate at the edges $k_-$ increases, a region near the edge where the piston is pushed towards the edge is created and grows. As the rate of particle passing through the piston $p$ increases, this region shrinks. For a low enough $k_-$ and high enough $p$, the piston is pushed from the edge to the center. All length scales are measured in units of $x_0\equiv2d$, all time scales are measured in units of $\tau_0 \equiv 200/\alpha$. $F_0\equiv\gamma x_0/\tau_0$.}
\label{fig:v0_x_wall_annihilation}
\end{figure}

We study two spatial patterns of particle sources and sinks. In the first, particles are uniformly created everywhere with a constant rate $k_+$ per unit length, with a random active force direction. Particles are annihilated at the two boundaries of the system with rate $k_-$ per particle. This choice is motivated for the vesicles in the oocyte \cite{Almonacid2015} by the existence of a large membrane reservoir at the cortex. First, we will find the steady state particle density and current in a system without a piston. The rate equations for the density of left and right moving particles $L(x,t)$ and $R(x,t)$ and the numbers of particles accumulated on the system boundaries at $x=\pm d$: \cite{Razin2017,Schnitzer1993,Tailleur2008,Tailleur2009}
\begin{equation} \label{eq:FP_bulk_c_wall_a_v_x_alpha_x_D_0}
\begin{array}{ll}
\vspace{2mm}
\partial_tR = - \partial_x(v(x)R) + \frac{\alpha(x)}{2}(L-R)+\frac{k_+}{2} \\
\vspace{2mm}
\partial_tL = \partial_x(v(x)L) + \frac{\alpha(x)}{2}(R-L)+\frac{k_+}{2} \\
\vspace{2mm}
\partial_t N_L^{-d} = -J_L(-d)-\frac{\alpha(-d)}{2} N_L^{-d} - k_- N_L^{-d} \\
\vspace{2mm}
\partial_t N_R^{-d} = -J_R(-d)+\frac{\alpha(-d)}{2} N_L^{-d} \\
\vspace{2mm}
\partial_t N_L^{d} = J_L(d)+\frac{\alpha(d)}{2} N_R^{d} \\
\vspace{2mm}
\partial_t N_R^{d} = J_R(d)-\frac{\alpha(d)}{2} N_R^{d} - k_- N_R^{d} \\
\end{array}
\end{equation}
where $J_R = v(x) R$ and $J_L = -v(x) L$ are the currents of right and left moving particles, and $N_{L/R}^x$ is the number of left/right-moving particles at the boundary position $x=\pm d$.
Note that the number of particles accumulated
on a wall that are moving away from it is zero (i.e. $N_R^{-d}=N_L^{d}=0$).
The resulting steady state density profile can be calculated analytically (see Appendix \ref{appendix:c_uniform_a_edge_sol}). For a velocity profile linearly increasing from the center to the edges, $v=a|x|+b$, the steady state density $\rho=R+L$ is
\begin{equation} \label{eq:FP_bulk_c_wall_a_v_x_Dr_x_D_0_sol_linear_v}
\begin{array}{ll}
\rho(x) = \frac{k_+}{a|x|+b} \left( d (1 + \frac{\alpha}{k_-}) - \frac{\alpha}{a} \left( |x| - d - \frac{b}{a}\log\big( \frac{a|x|+b}{ad+b} \big) \right) \right) \\
\\
N_R^d = N_L^{-d} = \frac{k_+ d}{k_-}
\end{array}
\end{equation}
We find that the particle density is proportional to $k_+$, while the shape of the density profile depends on $k_-$, as shown in Fig.~\ref{fig:v0_x_wall_annihilation}a. As $k_-$ is increased, the density does not approach a flat profile, and the density difference between the center and boundaries even increases.
The current of particles is towards the boundaries: $J(x)=k_+ x$ (Fig.~\ref{fig:v0_x_wall_annihilation}b).

Next, we consider the same system with a permeable piston with width $w_p$ and center at position $x_p$. Each of the two edges of the piston is a hard wall. Particles accumulated on each edge pass to the other edge with rate $p$. The equations describing this system and their steady state solutions for a system with an impermeable or a permeable piston appear in Appendix \ref{appendix:c_uniform_a_edge_sol}. From the steady state density of particles, we calculate the force they apply to the piston.
The force on each of the piston edges is equal to the number of particles accumulated on that edge, multiplied by the force a single particle applies: $\gamma v$, where $\gamma$ is a friction coefficient.  The total force on the piston is the sum of the forces on each of its edges:
\begin{equation} \label{eq:F_p}
F_p = \gamma \left( N_R^{x_p^l} v(x_p^l) - N_L^{x_p^r} v(x_p^r) \right)
\end{equation}
where $x_p^l=x_p-w_p/2$ is the position of the left edge of the piston, and $x_p^r=x_p+w_p/2$ is the position of the right edge of the piston. The final result for the particle density and hence for the force is too long to include here. Details of the calculation appear in Appendix \ref{appendix:c_uniform_a_edge_sol}, and the force on the piston is plotted in Fig.~\ref{fig:v0_x_wall_annihilation}c.

We find that increasing $k_-$, while increasing $k_+$ accordingly to keep their ratio constant, increases the particle flux towards the boundaries until a region near the boundaries appears where the force on the piston is outwards (Fig.~\ref{fig:v0_x_wall_annihilation}c). This region decreases in size as $p$ increases as shown in Fig.~\ref{fig:v0_x_wall_annihilation}d.

\subsection{Uniform particle creation and annihilation}
Next, we study the case of spatially uniform particle creation and annihilation, where each particle has a finite lifetime of $1/k_-$. The rate equations for the particle bulk density and boundary accumulations for a system without a piston are
\begin{equation} \label{eq:FP_ca}
\begin{array}{ll}
\vspace{2mm}
\partial_tL = \partial_x( v(x)L) + \frac{\alpha}{2}(R-L)+k_+ - k_- L \\
\vspace{2mm}
\partial_tR = - \partial_x (v(x)R) + \frac{\alpha}{2}(L-R)+k_+ - k_- R\\
\vspace{2mm}
\partial_t N_L^{-d} = -J_L(-d)-\frac{\alpha}{2} N_L^{-d} - k_- N_L^{-d} \\
\vspace{2mm}
\partial_t N_R^{-d} = -J_R(-d)+\frac{\alpha}{2} N_L^{-d}\\
\vspace{2mm}
\partial_t N_L^{d} = J_L(d)+\frac{\alpha}{2} N_R^{d}\\
\vspace{2mm}
\partial_t N_R^{d} = J_R(d)-\frac{\alpha}{2} N_R^{d} - k_- N_R^d \\
\end{array}
\end{equation}

The steady state density for the case of a linear velocity gradient $v=a|x|+b$ is given by (calculation in Appendix \ref{appendix:ca_uniform_sol})
\begin{equation} \label{eq:FP_ca_v0_x_rho_eq_x_ge_0_sol}
\rho(x) = c_1(ax+b)^{\lambda_+}+c_2(ax+b)^{\lambda_-} - \frac{k_+ (\alpha+k_-)}{a^2-k_-(\alpha+k_-)}
\end{equation}
where $\lambda_{\pm} = -1\pm \sqrt{\frac{k_-(2\alpha+k_-)}{a^2}}$, and $c_1, c_2$ are constants that depend on the parameters. This expression for $\rho$ is valid in $0\le x<d$. The density in $-d<x\le 0$ can be determined from it using the problem's reflection symmetry. The accumulations at the boundaries for this case are given in Appendix \ref{appendix:ca_uniform_sol}.

\begin{figure}[h]
\includegraphics[width=1\linewidth]{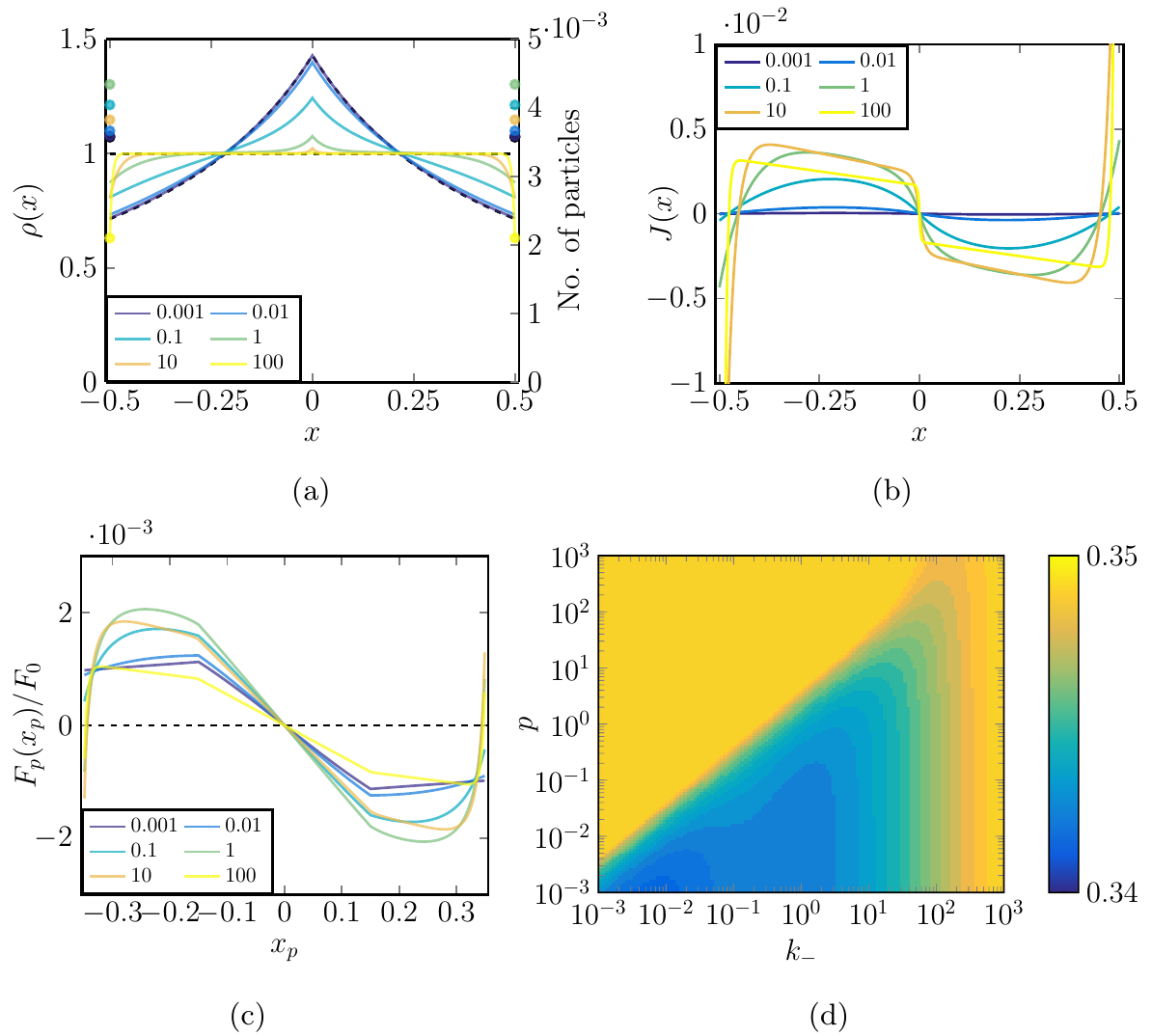}
  \caption[]{For a system with a constant $\alpha$ and $v(x)=|x|+0.5$ ($k_-$ is varied while setting $k_+=k_-/2d$ accordingly so that the total number of particles is constant and equal to $ N=2k_+d/k_-=1$ in a system without a piston.): (a) The particle density (lines) and the number of particles accumulated on edges (dots), for varying values of $k_-$ (Eq.~\ref{eq:FP_ca_v0_x_rho_eq_x_ge_0_sol}). Dashed black lines denote the limiting distributions: For $k_+,k_-\to \infty$, the distribution tends to uniform. For  $k_+,k_-\to 0$, the distribution tends to the one of the no-flux case (Fig.~\ref{fig:sys_plot_v_x_rho_p}b). (b) The particle current for varying values of $k_-$. ((a) and (b) - for a system without a piston) (c) The force on the piston for varying $k_-$ values ($w_p=0.3$, $p=1$) (d) The maximal position $0\le x_p\le d-w_p/2$ where the force on the piston is towards the center: $F_p(x_p)\le 0$ ($w_p=0.3$). For low enough $p$, and a regime of $k_-$ values, there is a region near the system edge where the force on the piston is towards the edge. For the parameters chosen, this region is always very small (less than $0.01$ wide). All length scales are measured in units of $x_0\equiv2d$, all time scales are measured in units of $\tau_0 \equiv 200/\alpha$. $F_0\equiv\gamma x_0/\tau_0$.}
\label{fig:v0_x_annihilation_everywhere}
\end{figure}

The density profile for various particle turnover rates is shown in Fig.~\ref{fig:v0_x_annihilation_everywhere}a. In the limit of $k_- \to \infty$, while keeping the ratio $k_+/k_-$ constant, the density approaches a uniform distribution: $\rho(x) \to k_+/k_-$, since particles are created uniformly and in this limit are immediately annihilated. As $k_-$ is increased, particle annihilation in the dense regions increases and the density becomes flatter. Particles are created uniformly, then move according to the $v(x)$ induced dynamics, typically leading them to regions where they are slow and dense, where they are annihilated. This creates particle currents in the system, shown in Fig.~\ref{fig:v0_x_annihilation_everywhere}b. We find that in most of the system the current is inwards, due to the annihilation of particles in the dense region at the center, while near the edges there is a region of outwards current due to the annihilation of particles accumulated at the edges.

As before, we calculate the mean force on a permeable piston (given by Eq.~\ref{eq:F_p}) from the steady state density in a system with the piston. The details of the calculation appear in Appendix \ref{appendix:ca_uniform_sol}. The force on a piston is towards the center, except for narrow regions near the edges, where the current is outwards (shown in Fig.~\ref{fig:v0_x_annihilation_everywhere}c). The extent of this region of outwards force has a non-monotonic dependence on $k_-$, shown in Fig.~\ref{fig:v0_x_annihilation_everywhere}d. Nevertheless, it always remains highly localized near the boundaries, and throughout most of the domain the force on the piston is towards the center.
Note that in the two limiting cases of $k_- \to 0$ and $k_- \to \infty$ while keeping $k_+/k_-$ constant, the current density in the system vanishes. In the no turnover limit, $k_- \to 0$, the system is closed and bounded and therefore the current vanishes. In the infinitely fast turnover limit, $k_- \to \infty$, the current vanishes since particles are created and immediately annihilated. Therefore a maximal current in the system is achieved at some intermediate $k_-$ value.

\section{Two dimensional model}
Following the insights we obtained from the one dimensional study, we simulated a system in two-dimensions, which is free of the pathologies of one dimension. The two dimensional simulation is also more similar to the experiments \cite{Almonacid2015}, where the motion is in three dimensions. In two dimensions we work with active Brownian particles, whose direction of motion diffuses with rate $D_r$, instead of run-and-tumble dynamics ($D_r$ has a role similar to $\alpha$ \cite{Cates2013ABPvsRTP,Solon2015ABPRTP}). In these simulations, a circular domain with radius $R$ contains active particles that push a rigid disk with radius $R_d$. $D_r$ is constant while $v(r)$ is a function of the distance from the system center.
Similarly to the 1D model, we consider systems with a uniform creation of particles, with a random active force direction, at rate $k_+$ per unit area. We consider two possible cases for particle annihilation at rate $k_-$: annihilation at the edge only, and a uniform annihilation rate (exponential lifetime).
For each case we use simulations to obtain the steady state particle density and current density in the absence of a disk, and the mean force on the disk when it is placed at different radial positions.

\emph{Simulation details ---} 
We simulated 2D systems of non-interacting active Brownian particles by numerically integrating the overdamped Langevin equation of motion for each of the particles, using the Euler method. The equation of motion for each particle is
\begin{align}  \label{eq:2d_Langevin_EOM}
\partial_t \vec{r} &= v \hat{n}_{\theta}(t) + \frac{1}{\gamma} \vec{F}^{ext} \\
\partial_t \theta &= \eta(t)
\end{align}

where  $v$ is the self propulsion speed, $\vec{r}=(x,y)$ is the particle's position, $\hat{n}_{\theta}=(\cos\theta,\sin\theta)$ is a unit vector in the direction of the motility force of the particle, and $\eta$ is white noise obeying $\langle \eta(t) \rangle = 0$ and $\langle \eta(t) \eta(t') \rangle = 2D_r \delta(t-t')$. $\gamma$ is the friction coefficient, and $\vec{F}^{ext}$ is the external force on the particle, due to interaction with the system boundaries and objects inside the system.

The system and disk boundaries apply on the particles a force derived from narrow Lennard-Jones potentials truncated at the minimum, leaving just the repulsive part:

\begin{align}
V(\vec{\Delta r}) &=
\begin{cases}
& 4 \epsilon \left( \left(\frac{\sigma}{\vec{\Delta r}}\right)^{12} - \left(\frac{\sigma}{\vec{\Delta r}}\right)^{6} \right) + \epsilon, \ \text{if}\ |\vec{\Delta r}| < 2^{1/6}\sigma \\
& 0, \text{otherwise}
\end{cases}
\end{align}

where $\vec{\Delta r} = \vec{r} - \vec{r_{wall}}$, with $\vec{r_{wall}}$ being the position of the point on the wall closest to the particle. 

In order to the determine the force on a disk at different radial positions, simulations were performed with a static disk held at each position. The mean force measured on such a disk is valid for a moving disk in the limit in which its velocity is small with respect to the active particle velocity \cite{Razin2017}.

\begin{figure}[h]
\includegraphics[width=1\linewidth]{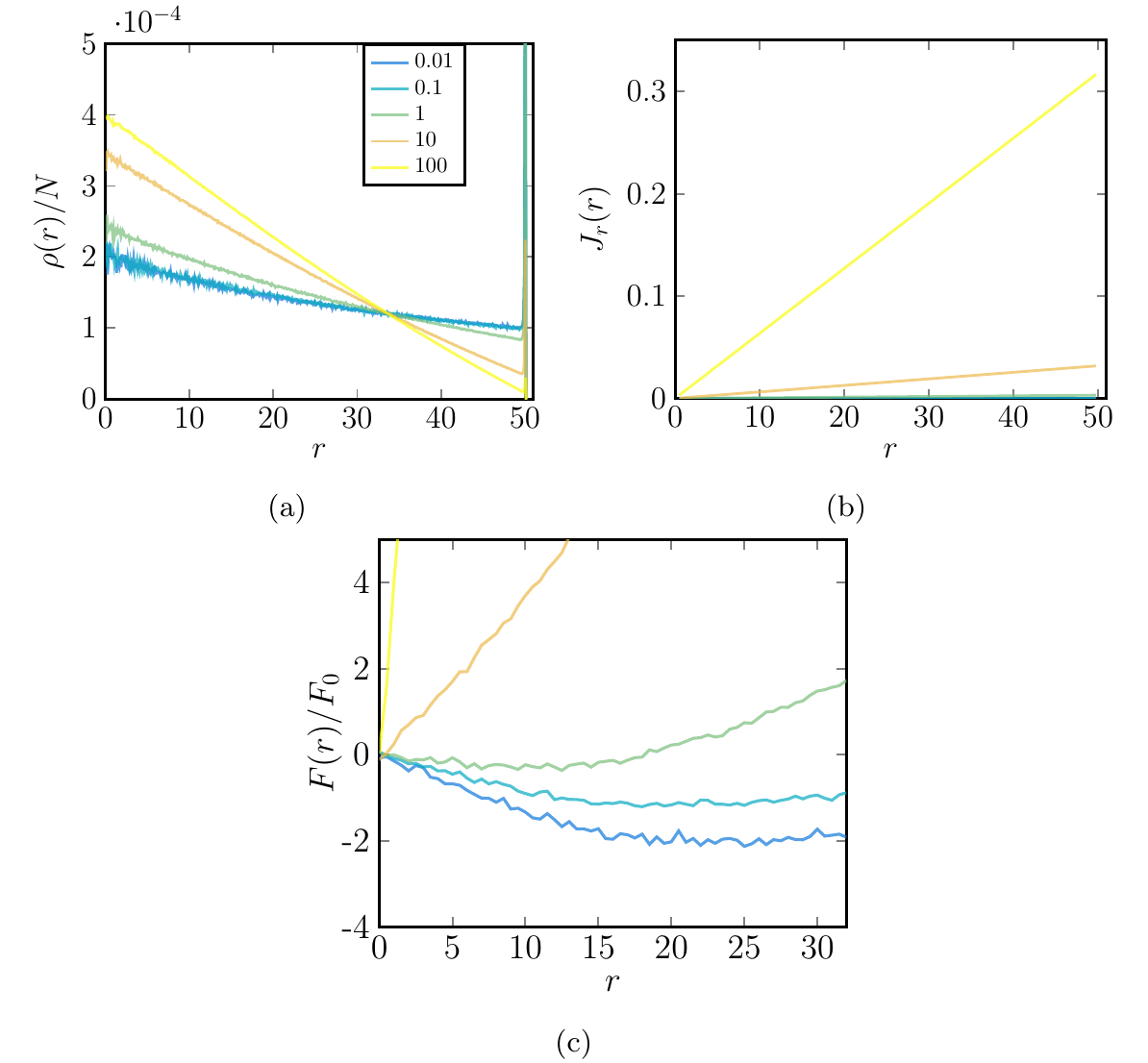}
  \caption[]{Simulation results for a 2D circular system, with particles created at rate $k_+$ per unit area uniformly, and annihilated at rate $k_-$ per particle at the system edge. For varying $k_-$ at constant $k_-/k_+=\pi R^2$, (a) the particle density, divided by the total number of particles, as a function of the distance from the system center $r$. (b) the particle current density in the $\hat{r}$ direction as a function of $r$. (c) The force on a disk with center at $r$ inside the system. All length scales are measured in units of $x_0\equiv R/50$, all time scales are measured in units of $\tau_0 \equiv 100/D_r$. $F_0\equiv\gamma x_0/\tau_0$. $v(r)=50(r/R+1)$, $R_d=15$.}
\label{fig:2D_ca_edge}
\end{figure}

\begin{figure}[h]
\includegraphics[width=1\linewidth]{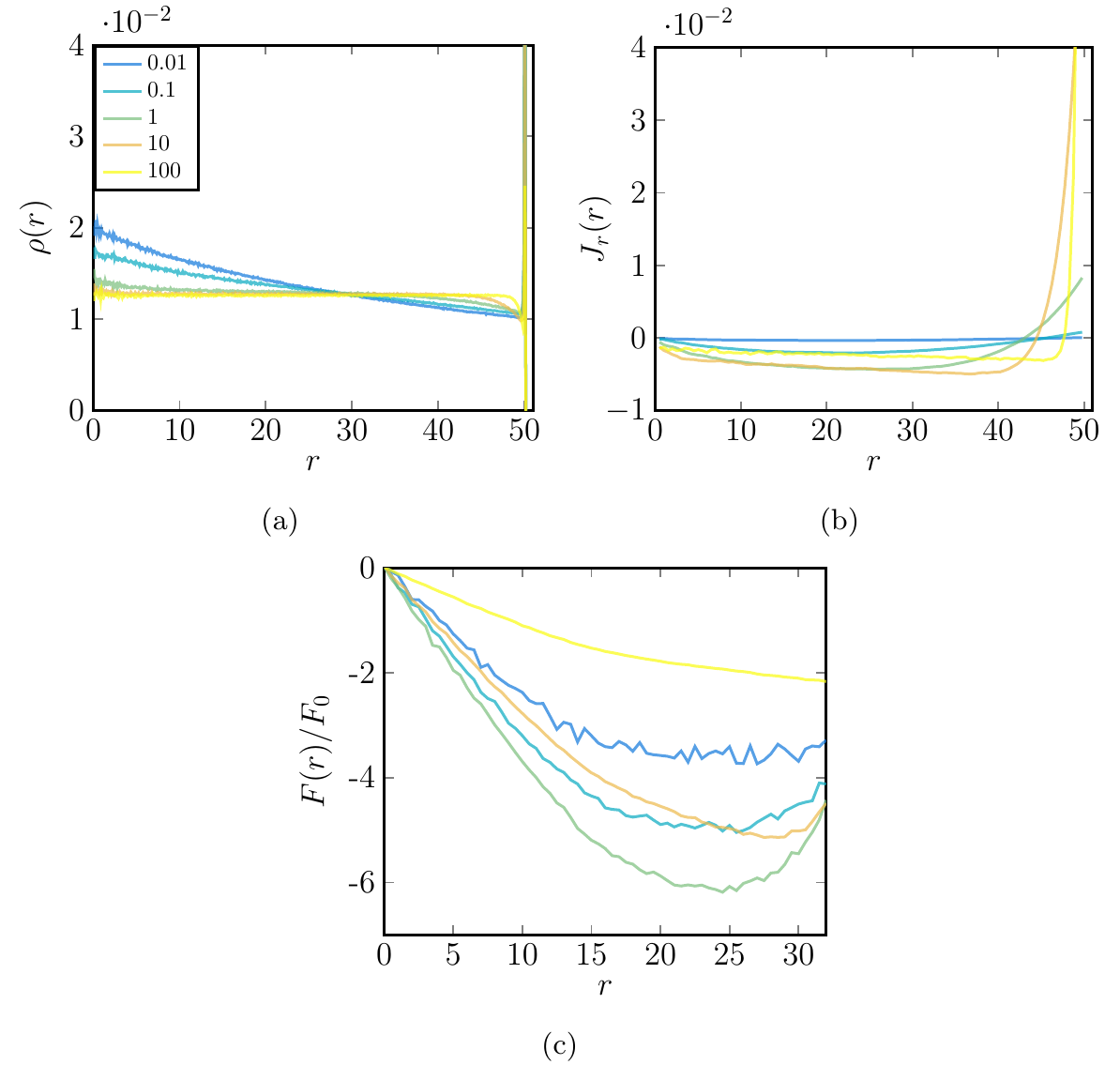}
  \caption[]{Simulation results for a 2D circular system, with particles created at rate $k_+$ per unit area uniformly, and annihilated at rate $k_-$ per particle. For varying $k_-$ at constant $k_+/k_-$ ($k_+/k_-=100/\pi R^2$), (a) the particle density as the function of the distance from the system center $r$. (b) the particle current density in the $\hat{r}$ direction as a function of $r$. (c) The force on a disk with center at $r$ inside the system. All length scales are measured in units of $x_0\equiv R/50$, all time scales are measured in units of $\tau_0 \equiv 100/D_r$. $F_0\equiv\gamma x_0/\tau_0$. $v(r)=50(r/R+1)$, $R_d=15$.}
\label{fig:2D_ca_everywhere}
\end{figure}

In the simulations of the two dimensional system presented in Fig.~\ref{fig:2D_ca_edge} and Fig.~\ref{fig:2D_ca_everywhere}, the parameters of the interaction potential of the system and disk boundaries with the active particles were $\epsilon=1 F_0 x_0$, $\sigma=0.25 x_0$.
The particle density, current density and force on an immersed disk shown are averages calculated from 10 simulations, each over a total simulation time of $10^4 \tau_0$. The simulation step size was $dt=10^{-4} \tau_0$.

\emph{Results ---} 
We work in the small persistence length limit ($\ell_p \ll R_d$). In \cite{Razin2017} we showed, in the absence of particle turnover, that in this regime the force on the disk is in the direction of minimal persistence length $\ell_p=v/D_r$ (as is always true for the piston in the 1D model), which is towards the center of the system when we choose a constant $D_r$ and a $v(r)$ that increases from the center to the edge.

We find that the particle density, current and force on the disk in the 2D simulations behave qualitatively similar to the results we obtained in the 1D system with a permeable piston. In the case of edge annihilation of the particles (Fig.~\ref{fig:2D_ca_edge}), the current density in the $\hat{r}$ direction in $d$ dimensions is $J_r(r)=\frac{k_+}{d}{r}$ (as derived in Appendix \ref{appendix:c_uniform_a_edge_flux}). Therefore increasing the particle creation rate increases the particle flux towards the system edges (Fig.~\ref{fig:2D_ca_edge}b), resulting in a less uniform density profile (Fig.~\ref{fig:2D_ca_edge}a). For small enough fluxes, the force on the disk is towards the system center. As the flux grows, a region near the edge where the force is outwards emerges, and eventually for a large enough flux, the force on the disk is towards the edge (Fig.~\ref{fig:2D_ca_edge}c).
Note that in order to qualitatively recreate the 2D results, it is necessary to use a permeable piston in the 1D model (compare Figs.~\ref{fig:v0_x_wall_annihilation} and \ref{fig:2D_ca_edge}): as seen in Fig.~\ref{fig:v0_x_wall_annihilation}d, p must be nonzero in order to have a force towards the center throughout the system for some range of small $k_-$ values.

For a system with uniform particle annihilation, increasing the particle creation and annihilation rates flattens the density, as expected (Fig.~\ref{fig:2D_ca_everywhere}a). A current towards the center is created in most of the system, except for a small region near the edge (Fig.~\ref{fig:2D_ca_everywhere}b). As creation and annihilation rates increase, this current reaches a maximum and decreases, and does not grow indefinitely as in the edge annihilation case. Therefore a force on the disk towards the center of the system is maintained (Fig.~\ref{fig:2D_ca_everywhere}c). Overall, the 1D and 2D systems behave qualitatively similar (compare Figs.~\ref{fig:v0_x_annihilation_everywhere} and \ref{fig:2D_ca_everywhere}).

\section{Comparison to oocyte experimental data}
\begin{figure}[h]
\includegraphics[width=0.8\linewidth]{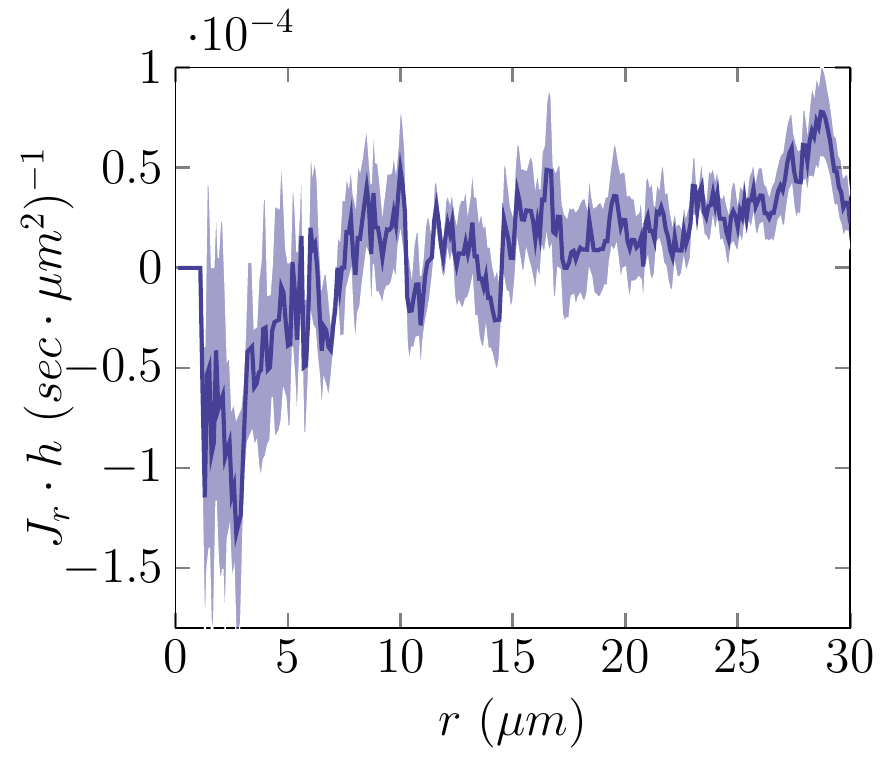}
  \caption[]{Vesicle current density (multiplied by the microscope's depth of field $h$)  in the $\hat{r}$ direction as a function of the distance from the cell center $r$, for the mouse oocytes studied in \cite{Almonacid2015}. Except for a peak near the cortex, the current density is close to zero within the noise. Error bars are SEM of data from $8$ experiments.}
\label{fig:exp_flux}
\end{figure}

In the oocyte, the measured radial current density of the active vesicles is very small throughout the volume, with a larger outwards current near the cortex (Fig.~\ref{fig:exp_flux}). This is in agreement with the previous observation \cite{Almonacid2015} that the number of vesicles traveling inwards and outwards balance each other throughout most of the oocyte volume, except near the cortex. Comparing with our model calculations for the case of particle annihilation at the edge (Fig.~\ref{fig:v0_x_wall_annihilation}, \ref{fig:2D_ca_edge}), and the case of uniform annihilation (Fig.~\ref{fig:v0_x_annihilation_everywhere}, \ref{fig:2D_ca_everywhere}), we conclude that the observed current density does not exactly fit either one of these simplified cases. It is possible that the vesicle flux in the oocyte is weak, therefore not substantially influencing the motion of the nucleus or the vesicle density distribution.

Since the experimentally measured current density is nearly zero within the noise level almost everywhere, another possibility is that there is uniform creation and annihilation with a fast enough vesicle turnover to cause a nearly uniform density, and a weak enough current density that can be within measurement noise of zero, with a positive peak near the cortex (compare Fig.~\ref{fig:2D_ca_everywhere}b and Fig.~\ref{fig:exp_flux}). Since the experimental system is three dimensional and our models one and two dimensional, it is difficult to predict from them when will a 3D system with uniform particle creation and annihilation have a current density that we can consider small. Therefore, we shall focus on determining if it is possible that in the experimental system the particle turnover is fast enough to create an approximately uniform density.
In order for the density to be approximately uniform, the particle lifetime $1/k_-$ needs to be short enough for the nonuniform velocity not to have a large effect. Thus the distance a particle covers in its lifetime $v/k_-$ needs to be be smaller than the distance over which the velocity substantially varies $v/\partial_xv$, i.e. we must demand $k_- \gg \partial_x v$. Replacing the derivative with a difference ratio that gives its typical value in the system, we estimate that the density will be approximately uniform when $\Delta v/\ell \ll k_-$, where $\Delta v$ is the velocity difference over the system and $\ell$ is the system size.
We can estimate this quantity for the oocyte vesicles: the difference in the measured velocity between the center and cortex is $\Delta v \approx 7\, \mu m/min$, the oocyte radius is $\ell \approx 35\, \mu m$. Hence the density can be approximately uniform if $\Delta v/\ell \approx 1/5\, min^{-1} \ll k_-$.
In the experiment, a cross section of the 3D cell was imaged. Vesicles enter and exit the nearly 2D field of view of the microscope as they move. Therefore the vesicle lifetime is larger than the average measured trajectory time, which is $\tau \approx 1\, min$. This gives an upper bound on the vesicle annihilation rate $k_- < 1 \,min^{-1}$. Hence we cannot rule out that $k_-$ is up to 5 times larger than $1/5\, min^{-1}$.

We conclude from the comparison of the density, current density and motion of the nucleus in the experiment \cite{Almonacid2015} to our model that it is possible that the vesicle turnover is negligible and the gradient in the vesicle activity is dominated by a gradient in the persistence time (as we suggested in \cite{Razin2017}). We cannot rule out the alternative option of a gradient in $v$, combined with a very fast uniform vesicle creation and annihilation. In order to determine whether the first option is correct, we suggest high frequency measurement of the vesicle trajectories, which would allow to determine whether microscopically the persistence time or the velocity is space dependent. To test the second option, we suggest measuring vesicle trajectories in 3D in order to better estimate their lifetime $1/k_-$.



\section{Conclusions}

We solved a 1D model for the force on an object (piston) inside a system of active particles with a gradient in their velocity, and  two types of source and sink patterns which are common in biological systems. A permeable piston was used in order to mimic the behavior of higher dimensional systems. We then showed in 2D simulations that the force on a large disk behaves similarly to the force on the permeable piston in our 1D model, despite the more complex geometry of the 2D system.

When active particles have a gradient in their intrinsic velocity, their density distribution is non-uniform \cite{Razin2017}. We showed that uniform particle sources and particle sinks at the system edge do not lead to a more uniform particle density in active systems with an outwards velocity increase. In addition, the outwards current created by this source and sink pattern causes the force on an object to become towards the system edge above a certain flux magnitude.
For the second source and sink pattern we considered, where particles are uniformly created and annihilated everywhere in the system, increased creation and annihilation rates bring the particle density closer to a uniform distribution, as expected. The force on an object remains towards the center throughout the system, except for a small edge region. 

While we currently do not have measurements of vesicle creation and annihilation patterns and rates, our model allows us to gain insight about which patterns could be consistent with the available experimental data, namely the density distribution. Our results could guide future experiments in trying to decipher the vesicle dynamics.
More generally, our results expand the understanding of active systems with an activity gradient, by considering different patterns of particles turnover. Such systems may be realized in synthetic dry active systems \cite{Lozano2016,Palacci2013,Buttinoni2013}, as well as shed light on active processes in biology \cite{Almonacid2015}.



\begin{acknowledgments}
We thank Marie-Helen Verlhac and Maria Almonacid for access to the experimental data. We thank Phil Pincus for helpful suggestions. NR thanks Ada Yonath and the Kimmelman center for financial support. NSG and NR thank the support of the Schmidt Minerva Center. NSG is the incumbent of the Lee and William Abramowitz Professorial Chair of Biophysics and this research was supported by the ISF (Grant No. 580/12). This research is made possible in part by the generosity of the Harold Perlman family.
\end{acknowledgments}

\appendix

\section{Equal accumulation on the edges of a 1D system with a symmetric permeable piston} \label{appendix:equal_accum}
Consider a 1D system with a permeable piston, where particles on each of the piston edges can cross to the other edge at rate $p$ per particle. 
Assume that a particle that crosses the piston appears immediately on the other side. It maintains its original active force orientation with probability $p_0$, and reorients to the inverse orientation with probability $1-p_0$.
We will show that the accumulation of particles on both sides are equal to each other and independent of the value of $p_0$.

The equations for the density of right and left moving particles in the bulk of the system are:
\begin{equation} \label{eq:FP_v_x_Dr_x}
\begin{array}{ll}
\vspace{1mm}
\partial_t R = -\partial_x(v(x)R) + \frac{\alpha(x)}{2}(L-R) \\
\vspace{1mm}
\partial_t L = \partial_x(v(x)L) + \frac{\alpha(x)}{2}(R-L) \\
\end{array}
\end{equation}
where $J_R = v(x) R$ and $J_L = -v(x) L$ are the currents of right and left moving particles.
The equations for the steady state number of particles on the left edge of the piston:
\begin{equation} \label{eq:FP_permeable_piston_left_general_reorientation}
\begin{array}{ll}
\partial_t N_L^{x_p^l} = J_L(x_p^l)+\frac{\alpha(x_p^l)}{2} N_R^{x_p^l}+p_0 p N_L^{x_p^r} = 0\\
\partial_t N_R^{x_p^l} = J_R(x_p^l)-\frac{\alpha(x_p^l)}{2} N_R^{x_p^l} - pN_R^{x_p^l} + (1-p_0) p N_L^{x_p^r} = 0
\end{array}
\end{equation}

where $N_{L/R}^x$ is the number of left/right-moving particles at position $x$, $x_p^l$ is the position of the left edge of the piston, and $x_p^r$ is the position of the right edge of the piston.
Summing the two equations and plugging in the bulk steady state  solution for $-d<x<x_p^l$, $L(x)=R(x)=\frac{1}{2}\rho(x)=\frac{c_1}{2v(x)}$ gives that $N_R^{x_p^l}=N_L^{x_p^r}$, i.e. the numbers of particles accumulated on the two edges of the piston are equal.

If the crossing rate of the particles from side to side is not symmetric, the numbers of particles accumulated on the two edges of the piston are different: Suppose that the rate of crossing the piston from left to right is $p_1$ per particle, while the rate of crossing in the opposite direction is $p_2$ per particle. For simplicity, assume a crossing particle maintains its active force direction. The equations for the steady state number of particles on the left edge of the piston are:
\begin{equation} \label{eq:FP_permeable_piston_left_asymmetric}
\begin{array}{ll}
\partial_t N_L^{x_p^l} = J_L(x_p^l)+\frac{\alpha(x_p^l)}{2} N_R^{x_p^l}+p_2 N_L^{x_p^r} = 0 \\
\partial_t N_R^{x_p^l} = J_R(x_p^l)-\frac{\alpha(x_p^l)}{2} N_R^{x_p^l} - p_1 N_R^{x_p^l} = 0
\end{array}
\end{equation}

From these, we obtain $N_L^{x_p^r} = \frac{p_1}{p_2} N_R^{x_p^l}$, i.e. ratio of numbers of particles accumulated on the two edges of the piston is equal to the ratio of the crossing rates.

\section{Solution of a 1D system with a symmetric permeable piston, in the absence of particle turnover} \label{appendix:1D_sym_pp_sol}
We will calculate the particle density in a 1D system with a symmetric permeable piston, in the absence of particle sources and sinks. The equations for the particle density to the left of the piston - for the bulk density in $-d<x<x_p^l$, the particle accumulation on the wall at $x=-d$ and the particle accumulation on the piston at $x=x_p^l$:

\begin{equation} \label{eq:FP_permeable_piston_left}
\begin{array}{ll}
-d<x<x_p^l:\\
\partial_t R = -\partial_x(v(x)R) + \frac{\alpha(x)}{2}(L-R) \quad;\quad J_R(x)=v(x)R(x) \\
\partial_t L = \partial_x(v(x)L) + \frac{\alpha(x)}{2}(R-L) \quad;\quad J_L(x)=-v(x)L(x)\\
\\
x=-d:\\
\partial_t N_L^{-d} = -J_L(-d)-\frac{\alpha(-d)}{2} N_L^{-d} \\
\partial_t N_R^{-d} = -J_R(-d)+\frac{\alpha(-d)}{2} N_L^{-d} \\
\\
x=x_p^l:\\
\partial_t N_L^{x_p^l} = J_L(x_p^l)+\frac{\alpha(x_p^l)}{2} N_R^{x_p^l}+pN_L^{x_p^r} \\
\partial_t N_R^{x_p^l} = J_R(x_p^l)-\frac{\alpha(x_p^l)}{2} N_R^{x_p^l} - pN_R^{x_p^l}
\end{array}
\end{equation}

In steady state, the left hand side of Eq.~\ref{eq:FP_permeable_piston_left} vanishes. Note that the steady state equations for the number of particles accumulated on the edges give both relations between the number of accumulated particles and the adjacent bulk density, and relations between the bulk densities $R$ and $L$ themselves, which serve as boundary conditions for the bulk density equations.
The steady state solution of these equations:
\begin{equation} \label{eq:permeable_piston_sol_left}
\begin{array}{ll}
-d<x<x_p^l:\\
L(x) = R(x) = \frac{1}{2}\rho(x) = \frac{c_1}{2v(x)}\\
N_L^{-d}=\frac{c_1}{\alpha(-d)}\\
N_R^{x_p^l} = \frac{c_1}{(\alpha(x_p^l)+2p)}\\
N_L^{x_p^r} = \frac{c_1}{(\alpha(x_p^l)+2p)}
\end{array}
\end{equation}

Similarly, the equations for the densities on the right side of the piston give:
\begin{equation} \label{eq:permeable_piston_sol_right}
\begin{array}{ll}
x_p^r<x<d:\\
L(x) = R(x) = \frac{1}{2} \rho(x) = \frac{c_2}{2v(x)}\\
N_R^{d}=\frac{c_2}{\alpha(d)}\\
N_L^{x_p^r} = \frac{c_2}{(\alpha(x_p^r)+2p)}\\
N_R^{x_p^l} = \frac{c_2}{(\alpha(x_p^r)+2p)}
\end{array}
\end{equation}

We have expressions for $N_R^{x_p^l}$ and $N_L^{x_p^r}$ from the equations on both sides. Compare them to get a relation between the constants $c_1$ and $c_2$:

\begin{equation}
c_2=\frac{\alpha(x_p^r)+p}{\alpha(x_p^l)+p}c_1
\end{equation}

Thus the equal accumulation on the two edges of the piston is $N_R^{x_p^l} = N_L^{x_p^r} = \frac{c_1}{(\alpha(x_p^l)+2p)}$.

$c_1$ is determined by the normalization condition:
\begin{equation}
\begin{split}
N & = \int_{-d}^{x_p^l} \rho(x)\mathrm{d}x + N_L^{-d} + N_R^{x_p^l} +  \int_{x_p^r}^{d} \rho(x)\mathrm{d}x + N_L^{x_p^r} + N_R^{d} \\
& = c_1\Big[ \int_{-d}^{x_p^l} \frac{\mathrm{d}x}{v(x)} + \frac{1}{\alpha(-d)} + \frac{1}{\alpha(x_p^l)/2+p}\\
& \hspace{3 mm} +\frac{\alpha(x_p^r)+2p}{\alpha(x_p^l)+2p} \Big( \int_{x_p^r}^{d} \frac{\mathrm{d}x}{v(x)} + \frac{1}{ \alpha(d)} \Big) \Big]\\
\end{split}
\end{equation}
where $N$ is the total particle number. The particle density for the case of a constant $\alpha$ and $v(x)$ that increases linearly from the system center to the system edge is plotted in Fig.~\ref{fig:sys_plot_v_x_rho_p}b.


The force on each of the piston edges is the number of particles accumulated on the edge multiplied by the force a single particle applies.  The total force on the piston is the sum of the forces on each of its edges:
\begin{equation}
\begin{split}
F_p & = N_R^{x_p^l}\gamma v(x_p^l) - N_L^{x_p^r}\gamma v(x_p^r)\\
& = \frac{c_1 \gamma}{(\alpha(x_p^l)+2p)} \Big( v(x_p^l) - v(x_p^r) \Big)
\end{split}
\end{equation}

where $\gamma$ is a friction coefficient. While the number of particles accumulated on a surface is generally a function of the local $\alpha$, the symmetrically permeable  piston equates the number of particles accumulated on each of its sides. Thus the force on the piston is proportional to the velocity difference between its edges, and unaffected by gradients in $\alpha$.

Note that for $\alpha = \text{const}$, taking $p=0$ in the results above gives exactly the same force as an impermeable piston when choosing to divide the particles between the two parts of the system according to the demand $c_1=c_2$.

\section{Solution of 1D model with uniform particle creation, and particle annihilation at the boundaries} \label{appendix:c_uniform_a_edge_sol}
We study a 1D model system with uniform particle creation and particle annihilation at the boundaries. We first calculate the steady state particle density and current density, in a system containing the active particles alone. Next, we study the same system, with the addition of a piston inside. We find the average force the active particles exert on the piston, by first solving for the steady state particle density in the system with the piston. While in this paper we focus on active particles with a constant tumble rate $\alpha$ and a spatially varying velocity $v(x)$, we keep a possibly spatially varying $\alpha(x)$ in the calculations for generality whenever it is possible without much complication.

\subsection{without a piston}
The equations for the particle density in the case of uniform particle creation in the system at rate $k_+$ with a random direction of motion, and particle annihilation at the edges at rate $k_-$ per particle:

\begin{equation} \label{eq:FP_bulk_c_wall_a_v_x_alpha_x_D_0_A}
\begin{array}{ll}
-d<x<d:\\
\partial_tR = - \partial_x(v(x)R) + \frac{\alpha(x)}{2}(L-R)+\frac{k_+}{2} \\
\\
\partial_tL = \partial_x(v(x)L) + \frac{\alpha(x)}{2}(R-L)+\frac{k_+}{2} \\
\\
x=-d:\\
\partial_t N_L^{-d} = -J_L(-d)-\frac{\alpha(-d)}{2} N_L^{-d} - k_- N_L^{-d} \\
\\
\partial_t N_R^{-d} = -J_R(-d)+\frac{\alpha(-d)}{2} N_L^{-d} \\
\\
x=d:\\
\partial_t N_L^{d} = J_L(d)+\frac{\alpha(d)}{2} N_R^{d} \\
\\
\partial_t N_R^{d} = J_R(d)-\frac{\alpha(d)}{2} N_R^{d} - k_- N_R^{d} \\
\end{array}
\end{equation}

The general steady state solution for $\rho(x)\equiv R+L$ and $\sigma(x)\equiv R-L$ is:

\begin{equation} \label{eq:FP_bulk_c_wall_a_v_x_Dr_x_D_0_sol}
\begin{array}{ll}
-d<x<d:\\
\sigma(x) = \frac{k_+ x + c_1}{v(x)} \\
\\
\rho(x) = \frac{1}{v(x)} \big( -\int_{-d}^{x} \frac{\alpha(x)}{v(x)} (k_+ x + c_1)\mathrm{d}x + c_2 \big) \\
\end{array}
\end{equation}

The equations for the steady state number of particles accumulated on the walls give the following boundary conditions for the bulk density:

\begin{equation} \label{eq:bulk_c_wall_a_v_x_Dr_x_D_0_bulk_bc}
\begin{array}{ll}
k_- \rho(-d) + (\alpha(-d)+k_-)\sigma(-d)=0 \\
k_- \rho(d) - (\alpha(d)+k_-)\sigma(d)=0 \\
\end{array}
\end{equation}

Also, they give an expression for the wall accumulation as a function of the bulk density next to the wall:

\begin{equation} \label{eq:bulk_c_wall_a_v_x_Dr_x_D_0_wall_accumulation}
\begin{array}{ll}
N_L^{-d} = 2\frac{v(-d)}{\alpha(-d)} R(-d) \\
N_R^{d} = 2\frac{v(d)}{\alpha(d)} L(d) \\
\end{array}
\end{equation}

Assume the problem has reflection symmetry, i.e. $\rho(x) = \rho(-x)$ and $\sigma(-x)=-\sigma(x)$, then 
from Eq.~\ref{eq:FP_bulk_c_wall_a_v_x_Dr_x_D_0_sol}, $c_1=0$, and the two b.c. equations are the same and yield $c_2=k_+d(1+\frac{\alpha(-d)}{k_-})$. Hence the solutions for $\rho$ and $\sigma$ become:

\begin{equation} \label{eq:FP_bulk_c_wall_a_v_x_alpha_x_D_0_sol_ref_sym}
\begin{array}{ll}
-d<x<d:\\
\sigma(x) = \frac{k_+ x}{v(x)} \\
\\
\rho(x) = \frac{k_+}{v(x)} \left( d (1 + \frac{\alpha(-d)}{k_-}) - \int_{-d}^{x} \frac{x\alpha(x)}{v(x)}\mathrm{d}x \right) \\
\\
N_L^{-d}=N_R^d=\frac{k_+ d}{k_-}
\end{array}
\end{equation}


For the case of $v(x)=a|x|+b$, $\alpha(x)=\alpha=const$, the solution is:
\begin{equation} \label{eq:FP_bulk_c_wall_a_v_x_Dr_x_D_0_sol_linear_v_APPENDIX}
\begin{array}{ll}
-d<x<d:\\
\sigma(x) = \frac{k_+ x}{a|x|+b} \\
\\
\rho(x) = \frac{k_+}{a|x|+b} \left( d (1 + \frac{\alpha}{k_-}) - \frac{\alpha}{a} \left( |x| - d - \frac{b}{a}\log\big( \frac{a|x|+b}{ad+b} \big) \right) \right)\\
\\
N_R^d = N_L^{-d} = \frac{k_+ d}{k_-}
\end{array}
\end{equation}
The total particle number, given by the integration over the density above, depends on both $k_-$ and $k_+$. The density is plotted in Fig.~\ref{fig:2D_ca_edge}a.

Summing the two equations for the density of right and left moving particles (Eq.~\ref{eq:FP_bulk_c_wall_a_v_x_alpha_x_D_0_A}), we obtain the following equation for the total particle density:
$\partial_t \rho = -\partial_x(v(x)\sigma)+k_+$. This is a continuity equation of the form $\partial_t \rho = -\partial_xJ+\mbox{source term}$, where $J$ is the current density. From this equation we find that the steady state current density is $J(x)=v(x)\sigma(x) = k_+ x$ (Fig.~\ref{fig:2D_ca_edge}b). Note it depends only on the particle creation rate $k_+$ and is independent of the annihilation rate per particle $k_-$.

\subsection{with an impermeable piston}
The equations for the particle density on the left side of the piston:
\begin{equation} \label{eq:FP_bulk_c_wall_a_D_0_v0_x_alpha_x_piston}
\begin{array}{ll}
-d<x<x_p^l:\\
\partial_tL = \partial_x(v(x)L) + \frac{\alpha(x)}{2}(R-L)+\frac{k_+}{2} \\
\\
\partial_tR = - \partial_x(v(x)R) + \frac{\alpha(x)}{2}(L-R)+\frac{k_+}{2} \\
\\
x=-d:\\
\partial_t N_L^{-d} = -J_L(-d)-\frac{\alpha(-d)}{2} N_L^{-d} - k_- N_L^{-d} \\
\\
\partial_t N_R^{-d} = -J_R(-d)+\frac{\alpha(-d)}{2} N_L^{-d}\\
\\
x=x_p^l:\\
\partial_t N_L^{x_p^l} = J_L(x_p^l)+\frac{\alpha(x_p^l)}{2} N_R^{x_p^l} \\
\\
\partial_t N_R^{x_p^l} = J_R(x_p^l)-\frac{\alpha(x_p^l)}{2} N_R^{x_p^l} \\
\end{array}
\end{equation}

The solution:
\begin{equation} \label{eq:1d_ca_v0_x_alpha_x_piston_sol}
\begin{array}{ll}
-d<x<x_p^l:\\
L(x)= \frac{k_+}{v(x)} \Big[-f(x) - \frac{1}{2}x + \frac{1}{2}d + x_p^l + (x_p^l+d)\frac{\alpha(-d)}{2k_-} \Big] \\
\\
R(x)= \frac{k_+}{v(x)} \Big[-f(x) + \frac{1}{2}x+ \frac{1}{2}d + (x_p^l+d)\frac{\alpha(-d)}{2k_-} \Big] \\
\\
x=-d \\ x_p^l:\\
N_L^{-d} = \frac{2v(-d)}{\alpha(-d)}R(-d) = \frac{k_+ (x_p^l+d)}{k_-} \\
N_R^{x_p^l} = \frac{2v(x_p^l)}{\alpha(x_p^l)}R(x_p^l) = \frac{2k_+}{\alpha(x_p^l)} \Big[-f(x_p^l) + (x_p^l+d) \big(\frac{\alpha(-d)}{2k_-} + \frac{1}{2} \big) \Big] \\
\end{array}
\end{equation}

where
\begin{equation} \label{eq: 1d_ca_v0_x_Dr_x_piston_sol_def_F}
f(x) \equiv \int_{-d}^{x} \frac{(x'-x_p^l)\alpha(x')}{2v(x')}\mathrm{d}x' \\
\end{equation}

Similarly, we solve equations for the density of particles on the right side of the piston, and get:
\begin{equation} \label{eq: 1d_ca_v0_x_Dr_x_piston_sol_right_side}
\begin{array}{ll}
x_p^r<x<d:\\
L(x)= \frac{k_+}{v(x)} \Big[ -\bar{f}(x) + \bar{f}(d) - \frac{1}{2}x + \frac{1}{2}d + (d-x_p^r)\frac{\alpha(d)}{2k_-} \Big] \\
\\
R(x)= \frac{k_+}{v(x)} \Big[-\bar{f}(x) + \bar{f}(d)  + \frac{1}{2}x+\frac{1}{2}d - x_p^r\\
 + (d-x_p^r)\frac{\alpha(d)}{2k_-} \Big] \\
\\
x=x_p^r \text{ or } d:\\
N_L^{x_p^r} = \frac{2v(x_p^r)}{\alpha(x_p^r)}L(x_p^r) =  \frac{2k_+}{\alpha(x_p^r)} \Big[+\bar{f}(d) + (d-x_p^r)\big(\frac{\alpha(d)}{2k_-}+\frac{1}{2} \big) \Big] \\
N_R^{d} = \frac{2v(d)}{\alpha(d)}L(d) =  \frac{k_+(d-x_p^r)}{k_-}\\
\end{array}
\end{equation}

where
\begin{equation} \label{eq: 1d_ca_v0_x_alpha_x_piston_sol_def_F}
\bar{f}(x) \equiv \int_{x_p^r}^{x} \frac{(x'-x_p^r)\alpha(x')}{2v(x')}\mathrm{d}x' \\
\end{equation}

Therefore the force on the piston is:
\begin{equation} \label{eq: bulk_c_wall_a_v0_x_alpha_x_Fp}
\begin{array}{ll}
F_p &= \gamma \left( N_R^{x_p^l}v(x_p^l) - N_L^{x_p^r}v(x_p^r)\right)\\
    &= k_+ \gamma \Big[ \frac{2v(x_p^l)}{\alpha(x_p^l)} -f(x_p^l) + (d+x_p^l) \big(\frac{\alpha(-d)}{2k_-}+\frac{1}{2} \big) \Big) \\
   &\mspace{60mu}  - \frac{2v(x_p^r)}{\alpha(x_p^r)} \Big( \bar{f}(d) + (d-x_p^r) \big( \frac{\alpha(d)}{2k_-} + \frac{1}{2} \big) \Big) \Big]
\end{array}
\end{equation}

Note that unlike for a closed system, here the number of particles in each side of the piston is not conserved even when the piston is impermeable. However, a permeable piston will behave differently than an impermeable one, since an impermeable piston divides the system into two disconnected parts.

For the case of constant $v$ and $\alpha$, the equations are:

\begin{equation} \label{eq:FP_bulk_c_wall_a_D_0}
\begin{array}{ll}
-d<x<d:\\
\partial_tR = - v\partial_xR + \frac{\alpha}{2}(L-R)+\frac{k_+}{2} \\
\\
\partial_tL = v\partial_xL + \frac{\alpha}{2}(R-L)+\frac{k_+}{2} \\
\\
x=-d:\\
\partial_t N_L^{-d} = -J_L(-d)-\frac{\alpha}{2} N_L^{-d} - k_- N_L^{-d} \\
\\
\partial_t N_R^{-d} = -J_R(-d)+\frac{\alpha}{2} N_L^{-d} \\
\\
x=d:\\
\partial_t N_L^{d} = J_L(d)+\frac{\alpha}{2} N_R^{d}\\
\\
\partial_t N_R^{d} = J_R(d)-\frac{\alpha}{2} N_R^{d} - k_- N_R^{d} \\
\end{array}
\end{equation}

The solution:
\begin{equation} \label{eq:1d_ca_sol}
\begin{array}{ll}
-d<x<d:\\
L(x)=\frac{k_+}{2v^2}\big[ -\frac{\alpha}{2} x^2 -v x + d (\frac{\alpha}{2} d +v+\frac{\alpha v}{k_-})\big]\\
\\
R(x)=\frac{k_+}{2v^2}\big[ -\frac{\alpha}{2} x^2 +v x + d (\frac{\alpha}{2} d +v+\frac{\alpha v}{k_-})\big]\\
\\
\rho(x)=L(x)+R(x)=\frac{k_+}{v^2}\big[ -\frac{\alpha}{2} x^2 + d(\frac{\alpha}{2} d +v+\frac{\alpha v}{k_-}) \big]\\
\\
x=\pm d:\\
N_R^d=N_L^{-d}=\frac{k_+ d}{k_-}
\end{array}
\end{equation}

The steady state number of particles in this system is:
\begin{equation} \label{eq:1d_ca_N}
N=\frac{k_+}{v^2}\Big[\frac{2}{3}\alpha d^3 + 2v d^2 (1+\frac{\alpha}{k_-})\Big]+\frac{2k_+ d}{k_-}
\end{equation}

\subsection{with a symmetric permeable piston}
The bulk equations (true for $-d<x<x_p^l$, $x_p^r<x<d$):

\begin{equation} \label{eq:FP_bulk_c_wall_a_D_0_permeable_piston_bulk_eq}
\begin{array}{ll}
\partial_tL = \partial_x(v(x)L) + \frac{\alpha(x)}{2}(R-L)+\frac{k_+}{2} \\
\\
\partial_tR = - \partial_x(v(x)R) + \frac{\alpha(x)}{2}(L-R)+\frac{k_+}{2} \\
\end{array}
\end{equation}

The equations for the particle accumulation on edges in the left side of the piston:
\begin{equation} \label{eq:FP_bulk_c_wall_a_D_0_permeable_piston_walls_left}
\begin{array}{ll}
x=-d:\\
\partial_t N_L^{-d} = -J_L(-d)-\frac{\alpha(-d)}{2} N_L^{-d} - k_- N_L^{-d} \\
\\
\partial_t N_R^{-d} = -J_R(-d)+\frac{\alpha(-d)}{2} N_L^{-d} \quad; \quad N_R^{-d}=0\\
\\
x=x_p^l:\\
\partial_t N_L^{x_p^l} = J_L(x_p^l)+\frac{\alpha(x_p^l)}{2} N_R^{x_p^l} + pN_L^{x_p^r} \quad; \quad N_L^{x_p^l}=0\\
\\
\partial_t N_R^{x_p^l} = J_R(x_p^l)-\frac{\alpha(x_p^l)}{2} N_R^{x_p^l}-pN_R^{x_p^l} \\
\end{array}
\end{equation}

\begin{figure}[h]
\includegraphics[width=0.8\linewidth]{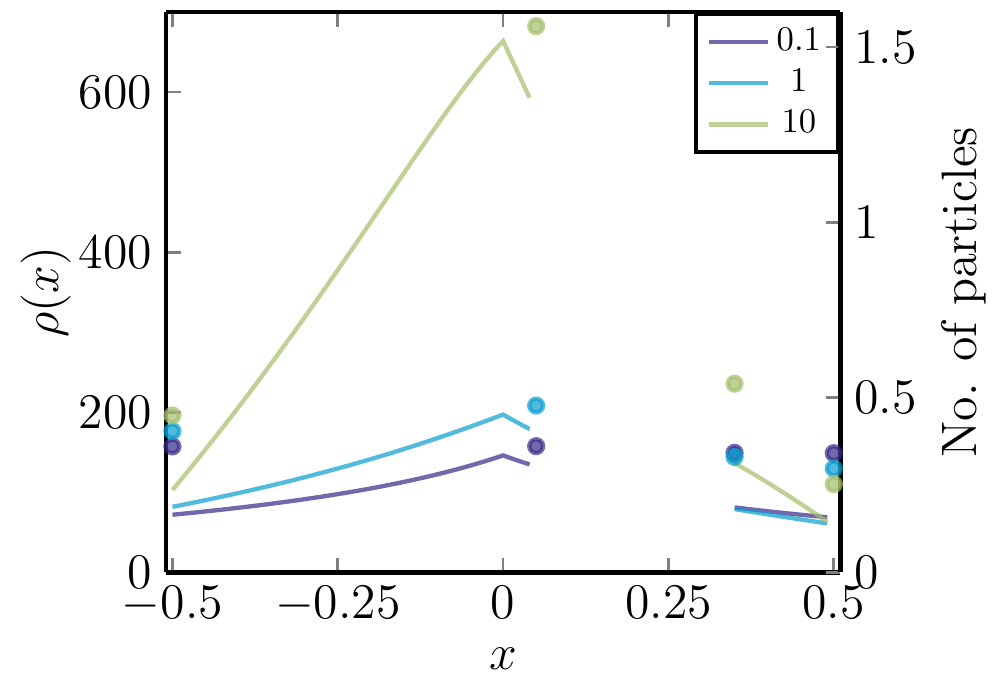}
  \caption[]{The steady state particle density (lines) and number of particles accumulated on edges (dots) for a 1D system with uniform particle creation, edge annihilation and a permeable piston with center position at $x_p=0.2$ and width $w_p=0.3$. Results are shown for varying turnover rates at a constant $k_-/k_+$ ratio: $k_-=0.1,1,10$ and $k_+=k_-/8d$. All other parameters are identical to the ones used for plotting the force on the piston in Fig.~\ref{fig:v0_x_wall_annihilation}c.}
\label{fig:edge_annihilation_permeable_piston_stst_density}
\end{figure}

Similarly, the equations for the particle accumulation on edges in the right side of the piston:
\begin{equation} \label{eq:FP_bulk_c_wall_a_D_0_permeable_piston_walls_right}
\begin{array}{ll}
x=x_p^r:\\
\partial_t N_L^{x_p^r} = -J_L(x_p^r)-\frac{\alpha(x_p^r)}{2} N_L^{x_p^r} - p N_L^{x_p^r} \\
\\
\partial_t N_R^{x_p^r} = -J_R(x_p^r)+\frac{\alpha(x_p^r)}{2} N_L^{x_p^r} + p N_R^{x_p^l} \quad; \quad N_R^{x_p^r}=0\\
\\
x=d:\\
\partial_t N_L^{d} = J_L(d)+\frac{\alpha(d)}{2} N_R^{d} \quad; \quad N_L^{d}=0\\
\\
\partial_t N_R^{d} = J_R(d)-\frac{\alpha(d)}{2} N_R^{d}- k_- N_R^{d} \\
\end{array}
\end{equation}
The general steady state solution of the bulk equations (\ref{eq:FP_bulk_c_wall_a_D_0_permeable_piston_bulk_eq}) is
\begin{equation} \label{eq:FP_bulk_c_wall_a_D_0_permeable_piston_bulk_sol}
\begin{array}{ll}
R(x)  = \frac{1}{v(x)} \Big[ -\int \frac{\alpha(x)(k_+ x + c_1)}{2v(x)}\mathrm{d}x + \frac{1}{2}(c_2+c_1+k_+ x) \Big] \\
\\
L(x)  = \frac{1}{v(x)} \Big[ -\int \frac{\alpha(x)(k_+ x + c_1)}{2v(x)}\mathrm{d}x + \frac{1}{2}(c_2-c_1-k_+ x) \Big]
\end{array}
\end{equation}
where $c_1$ and $c_2$ are integration constants. This solution applies in both the left and the right side of the piston, but with different integration constants, so we have 4 unknown integration constants.
In addition, we would like to find $N_L^{-d}$, $N_R^{x_p^l}$, $N_L^{x_p^r}$, $N_R^{d}$. Hence overall we have 8 unknowns, and 8 linear algebraic equations for them obtained by substituting the bulk solution into the equations for the steady state boundary accumulation equations (Eq.~\ref{eq:FP_bulk_c_wall_a_D_0_permeable_piston_walls_left} and \ref{eq:FP_bulk_c_wall_a_D_0_permeable_piston_walls_right}, with the left hand side equal to $0$). By solving these equations, we find the steady state density. Due to the length of the solution, it is not presented here. However, a plot of the solution for the steady state density appears in Fig.~\ref{fig:edge_annihilation_permeable_piston_stst_density}.

From the particle density, the force on the piston can be calculated as shown before, using Eq.~\ref{eq:F_p}.


In this system there is particle flux from the bulk, where particles are uniformly created, to the edge, where the particles exit the system. In order to get the piston to move to the center, gradients in $v$ and $\alpha$ must have a stronger effect than this outward flow. For a piston with zero width, there is no $v$, $\alpha$ difference between the two sides and the force on the piston is always outwards.

The force on the piston for $v=a|x|+b$ and a constant $\frac{k_+}{k_-}$  is plotted in Fig.~\ref{fig:v0_x_wall_annihilation}c of the main text. For a piston with nonzero width, the velocity difference can counteract the outwards flow of the particles to obtain an inwards force. The larger $k_-$ is, the larger the particle flux outwards. For small enough $k_-$, we get inwards force on the piston. For larger $k_-$, the force on the piston becomes negative in a region near the system edge, and positive from some point inwards.

\section{1D model with uniform particle creation and annihilation} \label{appendix:ca_uniform_sol}
We here study a 1D model system with uniform particle creation at rate $k_+$ per unit length, and uniform particle annihilation at rate $k_-$ per particle. The created particles have equal probabilities to have a right or left directed active velocity.

\subsection{without a piston}
The bulk equations ($-d<x<d$):

\begin{equation} \label{eq:FP_ca_D_0_bulk_eq_v0_x}
\begin{array}{ll}
\partial_tL = \partial_x( v(x)L) + \frac{\alpha}{2}(R-L)+\frac{k_+}{2} - k_- L \\
\\
\partial_tR = - \partial_x (v(x)R) + \frac{\alpha}{2}(L-R)+\frac{k_+}{2} - k_- R\\
\end{array}
\end{equation}

The equations for the accumulation of particles at the walls:
\begin{equation} \label{eq:FP_ca_D_0_wall_eq_v0_x}
\begin{array}{ll}
x=-d:\\
\partial_t N_L^{-d} = -J_L(-d)-\frac{\alpha}{2} N_L^{-d} - k_- N_L^{-d} \\
\\
\partial_t N_R^{-d} = -J_R(-d)+\frac{\alpha}{2} N_L^{-d}\\
\\
x=d:\\
\partial_t N_L^{d} = J_L(d)+\frac{\alpha}{2} N_R^{d}\\
\\
\partial_t N_R^{d} = J_R(d)-\frac{\alpha}{2} N_R^{d} - k_- N_R^d \\
\end{array}
\end{equation}

Summing and subtracting the two bulk equations, we get equations for the steady state $\rho\equiv R + L$ and $\sigma\equiv R - L$:

\begin{equation} \label{eq:FP_ca_D_0_bulk_eq_rho_sigma_v0_x}
\begin{array}{ll}
0 = -\partial_x (v(x)\sigma(x)) + k_+ - k_- \rho(x) \\
\\
0 = - \partial_x (v(x)\rho(x)) - (\alpha+k_-) \sigma(x) \\
\end{array}
\end{equation}

From the second equation,

\begin{equation} \label{eq:FP_ca_sigma}
 \sigma(x)=-\frac{1}{\alpha+k_-}\partial_x(v(x)\rho(x))
\end{equation}
 
 Plugging this into the first equation gives a second order differential equation for $\rho(x)$:

\begin{align} \label{eq:FP_ca_v0_x_rho_eq}
0 = & \big(v'(x)^2+v(x)v''(x) - k_-(\alpha+k_-) \big)\rho(x)\\
&+3v(x)v'(x)\rho'(x)+v(x)^2\rho''(x)+k_+(\alpha+k_-) \nonumber
\end{align}

In general, this equation is hard to solve. Hence we will consider a case which can be solved analytically - where $v(x)$ linearly increases from the center of the system towards the edges, i.e. $v(x) = a|x|+b$. Since our system is symmetric to reflection $x\to-x$ and replacement of $R \leftrightarrow L$, we can solve the problem in the half-system $0\le x\le d$ and use the symmetry to conclude what is the solution in the other half.

When we do this, the wall accumulation equations at $x=d$ give us a Dirichlet boundary condition on the values of $L$ and $R$. Since we now have a fake boundary at $x=0$, we need a boundary condition there. We use the condition $\sigma(x)=-\sigma(-x)$, which is true due to the reflection symmetry of the original problem.

To sum up, the problem we currently wish to solve:
The bulk equations for $0\le x\le d$ are Eq.~\ref{eq:FP_ca_D_0_bulk_eq_v0_x} and hence Eq.~\ref{eq:FP_ca_D_0_bulk_eq_rho_sigma_v0_x}, \ref{eq:FP_ca_v0_x_rho_eq} hold. Plugging $v(x) = ax+b$ into Eq.~\ref{eq:FP_ca_v0_x_rho_eq} we obtain:

\begin{align} \label{eq:FP_ca_v0_x_rho_eq_x_ge_0}
0 =& \big(a^2 - k_-(\alpha+k_-) \big)\rho(x)+3a(ax+b)\rho'(x)\\
&+(ax+b)^2\rho''(x)+k_+(\alpha+k_-) \nonumber
\end{align}

This is an Euler differential equation. Its general solution is:

\begin{equation} \label{eq:FP_ca_v0_x_rho_eq_x_ge_0_sol_APPENDIX}
\rho(x) = c_1(ax+b)^{\lambda_+}+c_2(ax+b)^{\lambda_-} - \frac{k_+ (\alpha+k_-)}{a^2-k_-(\alpha+k_-)}
\end{equation}

where $\lambda_{\pm} = -1\pm \sqrt{\frac{k_-(\alpha+k_-)}{a^2}}$, and $c_1, c_2$ are constants to be determined by the boundary conditions. In addition, we have to find $N_R^d$. We solve for these 3 unknowns using the three equations:

\begin{equation} \label{eq:ca_v0_x_rho_eq_x_ge_0_BC}
\begin{array}{ll}
\sigma(0) = 0 \\
\sigma(d) = \frac{k_-}{v(d)}N_R^d = \frac{k_-}{ad+b}N_R^d \\
N_R^d = \frac{2v(d)}{\alpha} L(d) = \frac{ad+b}{\alpha}(\rho(d)-\sigma(d))
\end{array}
\end{equation}

The first equation is true due to the reflection symmetry of the original system, and the last two equations are a result of the equations for $x=d$ in Eq.~\ref{eq:FP_ca_D_0_wall_eq_v0_x}. Using Eq.~\ref{eq:FP_ca_sigma} we find that 

\begin{align} \label{eq:FP_ca_sigma_sol}
 \sigma(x)&=-\frac{a}{\alpha+k_-} \Big[ c_1(\lambda_++1)(ax+b)^{\lambda_+}\\
 &+c_2(\lambda_-+1)(ax+b)^{\lambda_-} - \frac{k_+(\alpha + k_-)}{a^2 - k_-(\alpha+k_-)} \Big] \nonumber
\end{align}

Plugging Eq.~\ref{eq:FP_ca_v0_x_rho_eq_x_ge_0_sol_APPENDIX}, \ref{eq:FP_ca_sigma_sol}, into the conditions Eq.~\ref{eq:ca_v0_x_rho_eq_x_ge_0_BC}ת we obtain a system of equations for the unknowns $c_1,c_2,N_R^d$ which we can write in matrix form:

\begin{equation} \label{BC_matrix}
M = \left[\begin{smallmatrix}
    \frac{a}{\alpha+k_-}(\lambda_++1)b^{\lambda_+} & \frac{a}{\alpha+k_-}(\lambda_-+1)b^{\lambda_-} & 0 \\
    \frac{a}{\alpha+k_-}(\lambda_++1)(ad+b)^{\lambda_+} & \frac{a}{\alpha+k_-}(\lambda_-+1)(ad+b)^{\lambda_-} & \frac{k_-}{ad+b} \\
    \frac{1}{2}(ad+b)^{\lambda_+}\big(1+\sqrt{\frac{k_-}{\alpha+k_-}}\big) & \frac{1}{2}(ad+b)^{\lambda_-}\big(1-\sqrt{\frac{k_-}{\alpha+k_-}}\big) & -\frac{\alpha}{2(ad+b)} 
\end{smallmatrix} \right]
\end{equation}

\begin{equation} \label{BC_inhomog_vec}
V = \begin{bmatrix}
    \frac{a}{a^2-k_-(\alpha+k_-)}\\
    \frac{a}{a^2-k_-(\alpha+k_-)} \\
    \frac{1}{2}\frac{\alpha+k_-}{a^2-k_-(\alpha+k_-)} \big(1+\frac{a}{\alpha+k_-} \big)
\end{bmatrix}
\cdot k_+
\end{equation}

\begin{equation} \label{BC_matrix_form_eq}
M \begin{bmatrix}
    c_1\\
    c_2 \\
    N_R^d
\end{bmatrix}
 = V \Rightarrow \begin{bmatrix}
    c_1\\
    c_2 \\
    N_R^d
\end{bmatrix}
= M^{-1}V
\end{equation}

The solution of these equations gives the particle density, which is plotted in Fig.~\ref{fig:v0_x_annihilation_everywhere}a of the main text.
The shape of the density depends on $k_-$. In the limit of small $k_-$, the flux goes to zero and the density tends to the zero flux solution where $\rho \propto \frac{1}{v(x)}$. In the limit of large $k_-$, the distribution of particles is close to uniform. This is because particles are created uniformly in the system, and their lifetime $1/k_-$ is so short that they barely move before being annihilated. Thus the particle distribution reflects the distribution with which they are created, as opposed to the small $k_-$ limit where it reflects the distribution resulting from the spatial variability in $v(x)$.

Note there are two relevant timescales to compare with $1/k_-$ for finding the shape of $\rho(x)$: $1/\alpha$ and $d/v$ (Since $\rho(x) \sim k_+$, its shape doesn't depend on $k_+$).

Similarly to the edge annihilation case, summing the two equations for the density of right and left moving particles (Eq.~\ref{eq:FP_ca_D_0_bulk_eq_v0_x}), gives an equation for the total particle density:
$\partial_t \rho = -\partial_x(v(x)\sigma)+k_+-k_- \rho$. This is a continuity equation of the form $\partial_t \rho = -\partial_xJ+\mbox{source term}+\mbox{sink term}$, where $J$ is the current density. Thus the steady state current density is $J(x)=v(x)\sigma(x)$, with $\sigma(x)$ given by Eq.~\ref{eq:FP_ca_sigma_sol} with the constants $c_1$ and $c2$ given by Eq.~\ref{BC_matrix_form_eq}. This result is plotted in Fig.~\ref{fig:v0_x_annihilation_everywhere}b.

\subsection{with a permeable piston}
As before, assume $v(x)=a|x|+b$. Denote: $x_p$ is the position of the center of the piston; $w_p$ is the width of the piston; $x_p^r=x_p+w_p/2$ is the position of the right edge of the piston; $x_p^l=x_p-w_p/2$ is the position of the left edge of the piston.

Assume that the right side of the piston is in the right side of the system ($x_p^r>0$). First, assume that the left side of the piston is also in the right side of the system ($x_p^l>0$). We can now write equations for the density of particles in the system: bulk equations for the three regions I. $x_p^r<x<d$, II. $0<x<x_p^l$, III. $-d<x<0$. The bulk equations are Eq.~\ref{eq:FP_ca_D_0_bulk_eq_v0_x}. In regions I and II, $v(x)=ax+b$ and the general solution of the bulk equations is given by Eq.~\ref{eq:FP_ca_v0_x_rho_eq_x_ge_0_sol_APPENDIX}. Denote the two integration constants in region I by $c_1, c_2$, and in region II by $c_3,c_4$. In region III, $v(x)=-ax+b$. Plugging this into Eq.~\ref{eq:FP_ca_v0_x_rho_eq}, it can be shown that the general solution to the Euler equation for $\rho(x)$ there is:

\begin{equation} \label{eq:FP_ca_v0_x_rho_eq_x_le_0_sol}
\rho(x) = c_5(-ax+b)^{\lambda_+}+c_6(-ax+b)^{\lambda_-} - \frac{k_+ (\alpha+k_-)}{a^2-k_-(\alpha+k_-)}
\end{equation}

Remember $\sigma(x)$ in each region is found from $\rho(x)$ using Eq.~\ref{eq:FP_ca_sigma}. We thus find that in region I, 

\begin{equation} \label{eq:ca_x_ge_0_rho_m_rho_p}
\begin{array}{ll}
L(x) = c_1 A_+(ax+b)^{\lambda_+} + c_2 A_-(ax+b)^{\lambda_-}-B_+  \\
\\
R(x) = c_1 A_-(ax+b)^{\lambda_+} + c_2 A_+(ax+b)^{\lambda_-}-B_-
\end{array}
\end{equation}

where we denote: $A_{\pm}=\frac{1}{2}\big( 1\pm \sqrt{\frac{k_-}{\alpha+k_-}}\big)$, 
$B_{\pm}=\frac{k_+(\alpha+k_-\pm a)}{2(a^2-k_-(\alpha+k_-))}$.
In region II we have the same result except for replacing $c_1 \to c_3$, $c_2 \to c_4$. In region III, we similarly get:
\begin{equation} \label{eq:ca_x_le_0_rho_m_rho_p}
\begin{array}{ll}
L(x) = c_5 A_-(-ax+b)^{\lambda_+} + c_6 A_+(-ax+b)^{\lambda_-}-B_-  \\
\\
R(x) = c_5 A_+(-ax+b)^{\lambda_+} + c_6 A_-(-ax+b)^{\lambda_-}-B_+
\end{array}
\end{equation}

For each of the 4 edges in the system, we have two equations for the change in particle accumulation - one for right moving and one for left moving particles. Since the number of particles moving away from the edge which are accumulated on it is zero, we have 4 unknown numbers of particles on edges:
$N_R^d$ - the number of right moving particles accumulated at $x=d$,
$N_L^{x_p^r}$ - the number of left moving particles accumulated at $x=x_p^r$,
$N_R^{x_p^l}$ - the number of right moving particles accumulated at $x=x_p^l$,
$N_L^{-d}$ - the number of left moving particles accumulated at $x=-d$.
Together with the 6 integration constants in the general solutions for $\rho(x)$ in the three regions, we have 10 unknowns. The edge equations give us 8 conditions. The extra 2 conditions needed are continuity conditions at $x=0$, where we demand that all particle densities are continuous: $\rho(0^-)=\rho(0^+)$, $\sigma(0^-)=\sigma(0^+)$.

The 8 steady state equations for particle numbers on the edges are:
\begin{equation} \label{eq:ca_permeable_piston_edge_accumulation}
\begin{array}{ll}
0=\partial_t N_L^d = J_L(d) + \frac{\alpha}{2} N_R^d \\
0=\partial_t N_R^d = J_R(d) - (\frac{\alpha}{2}+k_-) N_R^d \\
0=\partial_t N_L^{x_p^r} = -J_L({x_p^r}) - (\frac{\alpha}{2}+k_-+p) N_L^{x_p^r} \\
0=\partial_t N_R^{x_p^r} = -J_R({x_p^r}) + \frac{\alpha}{2} N_L^{x_p^r} + p N_R^{x_p^l} \\
0=\partial_t N_L^{x_p^l} = J_L({x_p^l}) + \frac{\alpha}{2} N_R^{x_p^l} + p N_L^{x_p^r} \\
0=\partial_t N_R^{x_p^l} = J_R({x_p^l}) - (\frac{\alpha}{2}+k_-+p) N_R^{x_p^l} \\
0=\partial_t N_L^{-d} = -J_L(-d) - (\frac{\alpha}{2}+k_-) N_L^{-d} \\
0=\partial_t N_R^{-d} = -J_R(-d) - \frac{\alpha}{2} N_L^{-d} \\
\end{array}
\end{equation}

\begin{figure}
\includegraphics[width=0.8\linewidth]{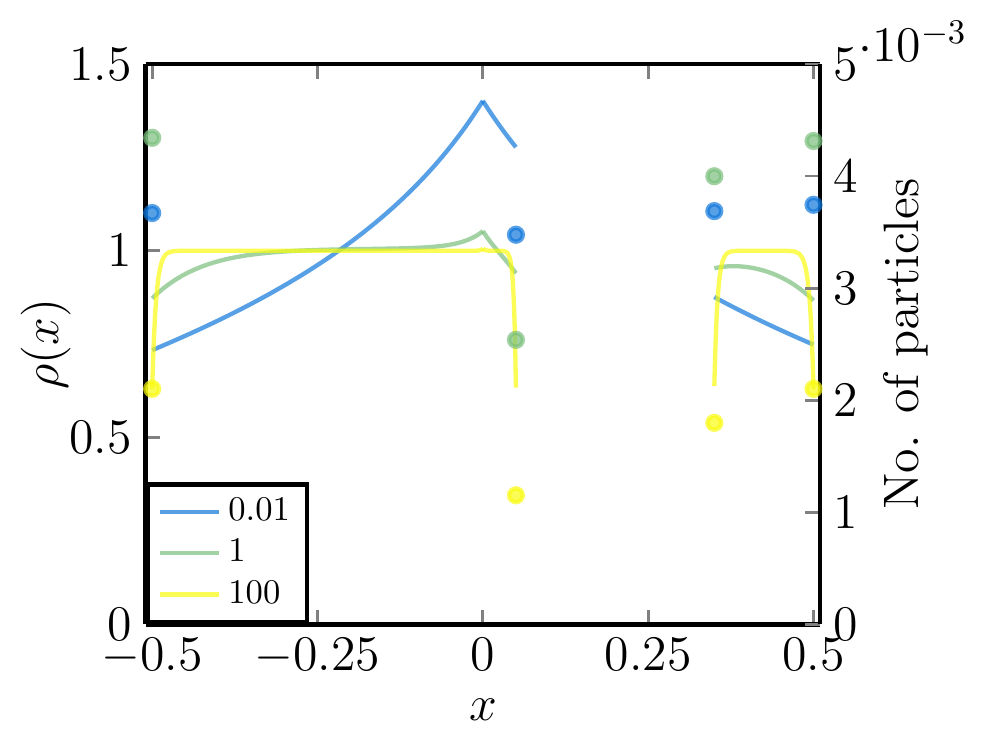}
\caption[]{The steady state particle density (lines) and number of particles accumulated on edges (dots) for a 1D system with uniform particle creation and annihilation, and a permeable piston with center position at $x_p=0.2$ and width $w_p=0.3$. Results are shown for varying turnover rates at a constant $k_-/k_+$ ratio: $k_-=0.01,1,100$ and $k_+=k_-/2d$. All other parameters are identical to the ones used for plotting the force on the piston in Fig.~\ref{fig:v0_x_annihilation_everywhere}c.}
\label{fig:uniform_annihilation_permeable_piston_stst_density}
\end{figure}

We can write the 10 linear equations on the 10 unknowns in matrix form, and invert the matrix to find the particle density in the system.
Similarly, we can solve the case $x_p^l<0$, which is even simpler because there are only 2 bulk regions and thus only 8 boundary conditions (no continuity equation needed for $x=0$, which in this case is inside the piston). 
The final result for the steady state density is plotted in Fig.~\ref{fig:uniform_annihilation_permeable_piston_stst_density}.
The resultant force on the piston for various $k_-$ values are plotted in the main text Fig.~\ref{fig:v0_x_annihilation_everywhere}c.

\section{The current in a d-dimensional spherical system with uniform particle creation and particle annihilation on the edge} \label{appendix:c_uniform_a_edge_flux}
For a spherically system with uniform particle creation, the particle density is given by:

\begin{equation} \label{eq:rho_uniform_creation}
\partial_t\rho(r) = -\nabla \cdot \vec{J}(r)+k_+
\end{equation}

where $k_+$ is the particle creation rate per unit volume, and $\vec{J}(r) = J_r(r)\hat{r}$ is the current density. Therefore $\nabla\cdot \vec{J} = \frac{1}{r^{d-1}}\partial_r \left( r^{d-1}J_r(r) \right)$. In steady state, the solution of Eq.~\ref{eq:rho_uniform_creation} is $J_r(r)=\frac{k_+}{d}r+\frac{c}{r^{d-1}}$, where $c$ is a constant. Due to the system's reflection symmetry, $c=0$ and therefore $J_r(r)=\frac{k_+}{d}r$.


%

\end{document}